# Design and development of the CSNS ion source control system[**]

LU Yan-Hua(卢艳华)[1,2]　　LI Gang(李刚)[1]　　OUYANG Hua-Fu(欧阳华甫)[1]

[1] Institute of High Energy Physics, CAS, Beijing 100049, China
[2] Graduate University of CAS, Beijing 100049, China

**Abstract:** After the CSNS ion source test stand has been stably working for years, an online control system for CSNS ion source aiming to be more stable and reliable is now under development. F3RP61-2L, a new PLC CPU module under Linux system, is introduced to the system as an IOC, to function together with the I/O modules of FA-M3 PLC on the PLC-bus. The adoption of the new IOC not only simplifies the architecture of the control system, but also improves the data transmission speed. In this paper, the design and development of the supervisory and control system for CSNS ion source are described.

**Key words:** control system, F3RP61, CSNS

**PACS:** 07.05.Dz　29.25.Lg

## 1. Introduction

The CSNS (China Spallation Neutron Source), under construction at IHEP (Institute of High Energy Physics), mainly consists of a $H^-$ linac and a proton rapid cycling synchrotron. The Penning Surface Plasma ion source, as the beginning of the $H^-$ linac, is used to first generate negative hydrogen ions ($H^-$). A stable, reliable and real time control system is essential to the successful commissioning of the ion source.

The control system for the CSNS ion test stand has been working successfully for years[1]. With this control system, the ion source test stand passed the expert appraisal in 2009. Fig. 1 shows the architecture of the control system for the ion source test stand. In this control system, an IOC (input output controller) monitors and controls the front end controller Sequence CPU through Ethernet. As Ethernet communication has uncertainty and unreliability, for instance, the loss of data or the delay of transmission, a PC under Linux as an IOC is unstable and non real-time. The control system needs some renovation to further improve its real-time and reliability.

Now, an online control system for the CSNS ion source is under construction on basis of the control system of the ion source test stand. This paper describes the design and development of the new control system.

## 2. Design of the new control system

In the design of the new control system, a new PLC (programmable logic controller) CPU named F3RP61 from Yokogawa is adopted. F3RP61, with an embedded real-time Linux operation system, can function as an IOC accessing PLC I/O modules directly or through the Sequence CPU.

The control system mainly consists of the F3RP61 (IOC-CPU), the traditional CPU (Seq-CPU) and PLC I/O modules to control the front-end devices, and will greatly simplify the control system architecture [2]. Fig. 2 shows the new control system architecture. In the system, each part functions as follows:

- IOC-CPU:

As a front-end IOC under Linux system, the EPICS (experimental physics and industrial control system) IOC application and some state transition programs can run directly in it. In addition, Linux kernel running in the IOC-CPU has a real-time preemption patch (PREEMPT_RT, details in Section 4). Compared with the soft-IOC used in the ion source test stand, the response time of this system is also more deterministic.

[*] Supported by Natural Science Foundation of China (11105166)
[*] Supported by Natural Science Foundation of China (91126003)

- Seq-CPU:

It has the original functions, which are interlock and to appoint its internal devices as stations to transfer data from or to the channels of I/O modules. Since the IOC-CPU can access the Seq-CPU's internal devices or channels of I/O modules with the shared memory, Seq-CPU is also used to appoint the shared memory devices as transfer stations for hardware channels.

- Workstation:

The workstations under Linux system are now used only as a NFS (net file system) and an OPI (operator interface). The NFS is used to store data and parameters of the control system. And the OPI's duty is to display the run-time variables of all signals.

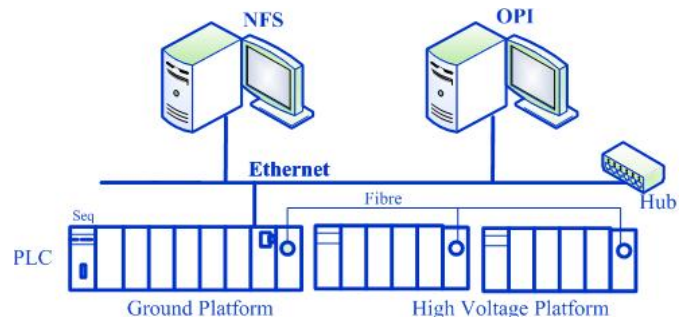

Fig. 1 Control system architecture for the CSNS ion source test stand

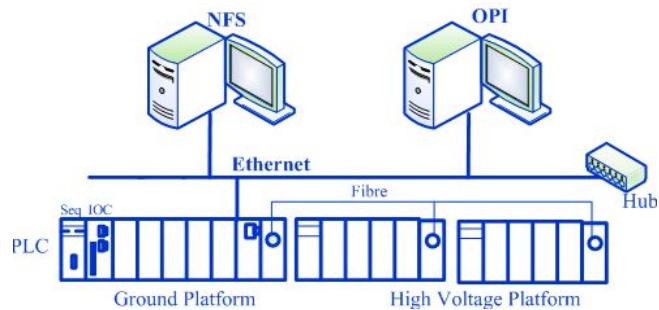

Fig. 2 New architecture of the control system for the CSNS ion source online system

## 3. Communication between IOC-CPU and Seq-PLC

When F3RP61 and Seq-CPU work on the same unit, there are three methods for them to communicate with each other[1]:

- Communication based on messages (asynchronous)
- Communication based on the shared memory ( synchronous)
- Communication based on Ethernet ( asynchronous)

---

[1] http://www-linac.kek.jp/cont/epics/f3rp61/DevSup_F3RP61-1.1.1.pdf

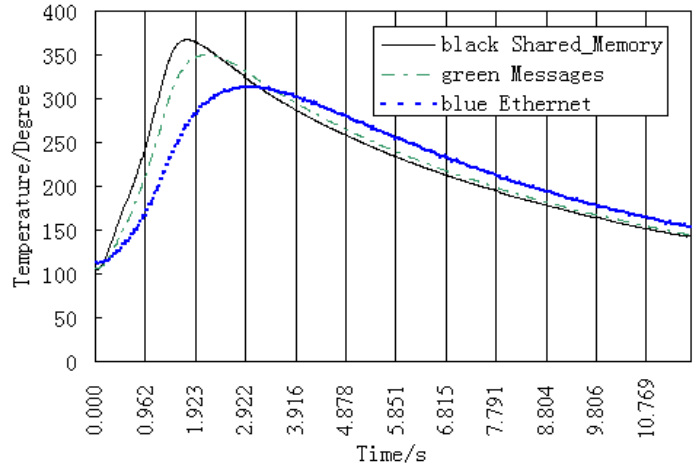

Fig. 3 The response curves of different communication methods

For the comparison of communication responses between different methods, a test was carried out. In the test, a transmitter was used to transmit the output of a K thermal coupler to 4~20mA, and to transfer its output to the PLC input channel. In the Seq-CPU, the input channel was appointed to a shared-memory device and an internal device. In the IOC-CPU, three different EPICS device and driver supports are used to support different communications. IOC application in the IOC-CPU reads data from the Seq-CPU's internal device and shared-memory device via the EPICS database records, then OPI reads the devices via channel access (EPICS communication protocol) and draws graphs or displays data.

Fig. 3 shows the response curves on the OPI graph. The black curve is based on shared-memory communication, the green one is based on message communication, while the blue one is based on Ethernet communication. Their peaks are separately allocated at (1.768s,367.277℃), (1.983s,348.892℃) and (3.0158s,312.972℃). As it can be seen, the shared-memory based communication is the fastest access, and the message based communication comes second, and the Ethernet based communication is the lowest.

In the ion source, temperature of the arc chamber rises sharply during its short circuit discharge. This demands a rapid monitor system to obtain the real-time temperature, and an interlock protection when the temperature exceeds a pre-set upper limit. The test shows that, communication based on the shared memory is more suitable than that based on Ethernet to be used in the online control system.

## 4. Latency effects of Linux real-time preemption patch (RT_Patch) on F3RP61[3]

4.1 Environments for CYCLICTEST

To investigate the latency effects of the RT_Patch, two different kernels (kernel with or without RT_Patch) running on the embedded Linux system are tested with the program CYCLICTEST. Table 1 shows the detailed features of the IOC-CPU.

Table 1 The F3RP61 system details for CYCLICTEST

| F3RP61 CPU | | MPC 8347E<br>533MHz(DDR: 133MHz; PCI: 33MHz; Surrounding: 66MHz) |
|---|---|---|
| **Memory** | Flash ROM | 64MB |
| | DDR SDRAM | 128MB |
| | SRAM | 512KB |
| | User SRAM | 4MB |
| **CF card** | | 8GB |
| **Linux kernel** | | 2.6.26.8 （With and without RT Patch） |

### 4.2 The CYCLICTEST results

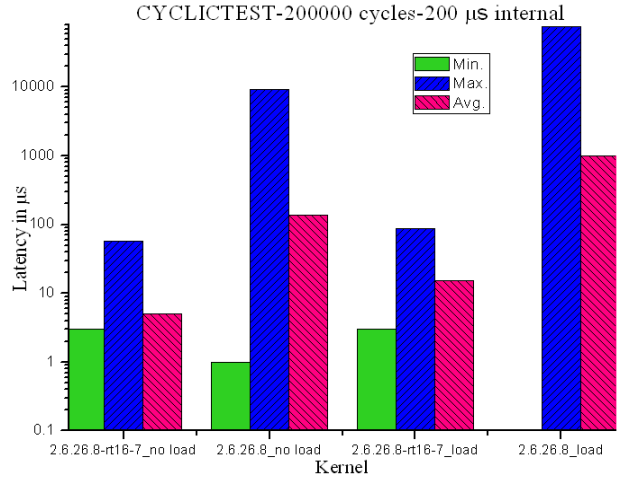

Fig. 4 The CYCLICTEST results

Table 2　Statistics of the CYCLICTEST results

| Kernels | 2.6.26.8-rt16-7 | 2.6.26.8-rt16-7 | 2.6.26.8 | 2.6.26.8 |
|---|---|---|---|---|
| Load? | NO | YES | NO | YES |
| Min.(μs) | 3 | 3 | 1 | 0 |
| Max.(μs) | 57 | 87 | 9349 | 74538 |
| Avg.(μs) | 5.01451 | 15.39705 | 138.07825 | 1004.62035 |

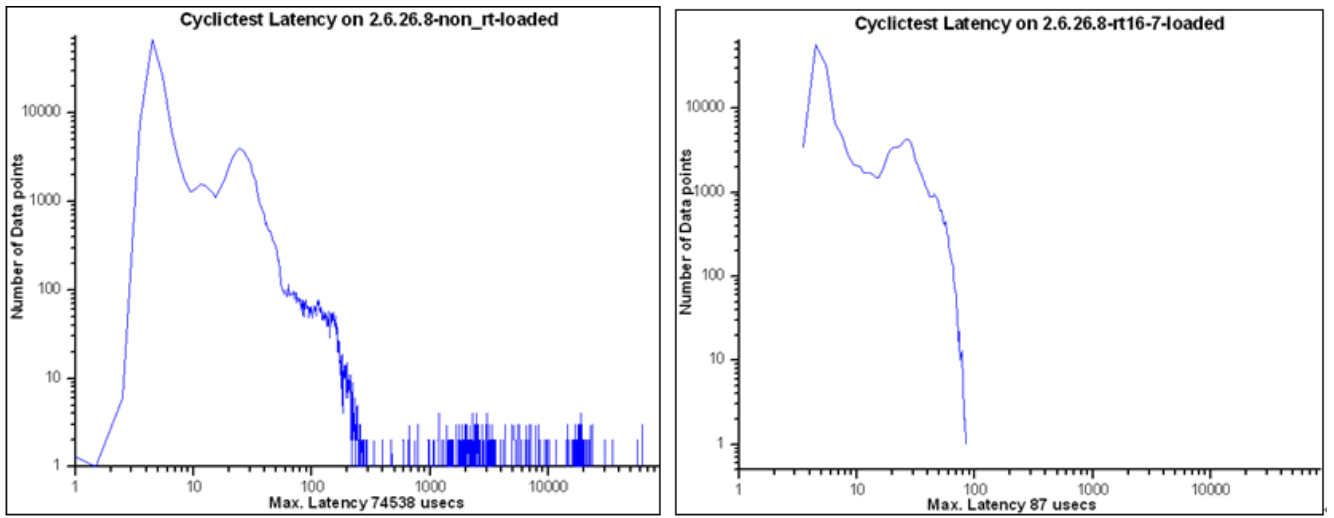

Fig. 5　The CYCLICTEST histogram latency

It can be seen from Fig. 4 and Table 2, the max. latency for kernel with RT_Patch is much lower than kernel without it. Fig. 5 shows the latency frequencies, from which it can be seen that the latency of the 200,000 loops is distributed in the range between 3μs-87μs on real-time kernel while the latency of many loops is distributed out of the range on non real-time kernel.

Real-time preemption patch does improve the preemption performance of Linux kernel. As it is known, real-time doesn't mean the fastest response, but implies response within a certain time. The definite response time is what our control system requires. Thus in the online control system for the CSNS ion source, the embedded Linux system in use has a kernel with real-time preemption patch.

### 5. Conclusion

Because the newly designed control system adopts F3RP61 as an IOC, the communication can be based on shared-memory or messages instead of Ethernet, which makes the system more reliable and response quickly. Linux system in the IOC-CPU has a real-time kernel, which overcomes the shortcoming of the unknowable response time in non real-time system. Now, tests and experiments of PLC have been completed. The online control system is to be built and tested recently.